# ON A CRUCIAL PROBLEM IN PROBABILITIES, AND SOLUTION

## Mioara MUGUR-SCHÄCHTER [*]

**Abstract**. First the crucial but very confidential fact is brought into evidence that – as Kolmogorov himself repeatedly claimed – there exists no abstract theory of probabilities, simply because the factual concept of probability is itself unachieved : it is nowhere specified how to construct the *factual* probability law to be asserted on a given *physical* random phenomenon.
Then an *algorithm of semantic integration* is built that permits to identify this factual probability law.

*This article is dedicated to Giuseppe Longo with whom I had several interesting exchanges concerning probabilities.*

## 1. INTRODUCTION

The concept of probability is ancient and intuitive. It belongs to common thinking and speaking. The mathematical formalisation of this concept has begun relatively late in the history of thought (Blaise Pascal 1654) and it evolved slowly (Bernoulli 1713, Richard von Mises 1931). The first thorough axiomatic and mathematical formalisation has been given by Andreï N. Kolmogorov [1950].

Meanwhile, in physics, Ludwig Boltzmann, long before Kolmogorov's work, has introduced his famous concept of *statistical* entropy (1872-1877) which rooted the second principle of thermodynamics into microphysics *via* the relative frequencies of outcome assigned to the considered 'events'.

Much later Shannon [1948] published his *theory of communication* (refined by Khincin [1957]) where Kolmogorov's abstract concept of a probability measure is made use of as a basic concept. Instead of 'events', Shannon introduced an 'alphabet' of *signs $\{a_i\}$, $i=1,2,....n$* on which he posited individual probabilities $p(a_i)$, $i=1,2,....n$ constituting a probability 'measure' in the sense of Kolmogorov's abstract theory of probabilities. Furthermore, Shannon defined, as a central concept of his theory, an entropic form called *informational entropy of the source of the signs $\{a_i\}$, $i=1,2,....n$* where in the place of Boltzmann's statistical relative frequencies of outcomes of physical events he introduced the corresponding abstract probabilities $p(a_i)$, $i=1,2,....n$.

For some time Shannon's concept of informational entropy seemed to permit the construction of entropic measures of complexity, thus leading to a mathematical theory of complexity. But, surprisingly, Kolmogorov [1963] – 30 years after his own construction of what was unanimously considered as an achieved modern mathematical theory of probability – became aware of the rather surprising fact that his mathematical representation of the concept of probability was *devoid of a factual basis* because the intuitive concept of probability itself is *a deficient concept*. In consequence of this realisation Kolmogorov began to claim that his theory of probabilities is not, as he had believed, an abstract reformulation of a well constructed physical concept, but exclusively an interesting mathematical construct. So he asserted that consequently his mathematical theory of probabilities cannot correctly found Shannon's theory of communication nor, *a fortiori*, a concept of informational entropy to be used for estimating complexities of factual entities. Therefore he initiated another approach for measuring complexities : the well known theory of "algorithmic" complexity of *sequences of signs*, which Chaitin and others keep developing.

But in the algorithmic representations of complexity, the semantic contents pointed toward by the considered sequences of signs, get entirely lost. They get lost to such a degree that still speaking of 'complexity' in these conditions partakes of derision and of deflection.

On the other hand in the recent studies of systems, of organisation, of constructive approaches, the accent falls more and more heavily upon the structures of *significances*. So far however the complexity qualifications of such structures stubbornly stay purely qualitative.

So, quite confidentially, the crucial correlated concepts of probability, information and complexity are undergoing a fundamental crisis in what concerns the representation of their semantic contents.

It is not current, nor easy, to both convey a fundamental problem and to propose a solution to it in only a couple of pages. However this is what is tried in what follows. We shall first define thoroughly the problem which obscures the concept of probability. Then we shall construct an effective solution to this problem, inside a general method of relativised conceptualisation of which the first principles have been drawn from the study of quantum probabilities (Mugur-Schächter [1991], [1992A], [1992B], [1993], [2008A], [2008B].

---

[*] http://www.mugur-schachter.net/ & Centre pour la Synthèse d'une Epistémologie Formalisée *(CeSEF)*, Paris, http://www.cesef.net/

## II. THE PROBLEM OF THE SPECIFICATION OF A *FACTUAL* PROBABILITY LAW

### II.1. Komogorov's Classical Definition of a Probability Space

The fundamental concept of the nowadays abstract theory of probabilities – in Kolmogorov's formulation – is a probability space *[U, τ, p(τ)]* where : $U=\{e_i\}$ (with $i \in I$ and $I$ an index set) is a *universe of elementary events* $e_i$ (a set) generated by the repetition of an 'identically' reproducible *procedure Π* (called also 'experiment') which, notwithstanding the posited identity between all its realisations, nevertheless brings forth elementary events $e_i$ that *vary* in general from one realisation of *Π* to another one ; *τ* is an *algebra of events* built on *U* [1], an event – let us denote it *e* – being a subset of *U* and being posited to have occurred each time that any elementary event $e_i$ from *e* has occurred ; *p(τ)* is a *probability measure* defined on the algebra of events *τ* [2].

A pair *(Π,U)* containing an identically reproducible procedure *Π* and the corresponding universe of elementary events *U* is called a *random phenomenon*.

On a given universe *U*, one can define various algebras *τ* of events. So it is possible to form different associations **[**[random phenomenon],[a corresponding probability space]**]**, all stemming from the same pair *(Π,U)*.

With respect to the previous representations of the concept of probability (Bernoulli, von Mises, etc.) – where only a 'probability law' (or 'probability measure' : there was no clear distinction as yet between factual and formal) was defined mathematically, Kolmogorov's concept of a probability *space [U, τ, p(τ)]*, introduced for the first time on the background of a radical distinction between factual data and formalised representation of these, has marked a huge progress : *via* this concept, the formal representation of the factual situations qualified as 'probabilistic' became inserted into the very elaborate mathematical syntax of measure theory, whereas before they had been only intuitively characterised, though also numbers were made use of.

### II.2. On the *interpretation* of an abstract probability measure

The probability measure *p(τ)* is the unique specifically probabilistic element from a Kolmogorov probability space. Now, to this very day, the application of this formal concept, to factual situations which are unanimously considered to be 'probabilistic', has not yet been founded upon an explicitly constructed concept of *factual* probability law. It is not even clearly known what significance one has to assign to the assertion that in this or that concrete probabilistic situation there 'exists' an empirical probability law. *A fortiori* it is not known how to identify that law. The specification – even in principle only – of such a significance-and-procedure, would suffice for installing a factual concept of probability law acceptable as an interpretation of the abstract concept of a formal probability measure, which would rehabilitate Kolmogorov's abstract representation in the status of a theory of probabilities. However the specification of such a significance-and-procedure is entirely lacking, which seems surprising concerning a concept so currently utilised and often playing such a basically important role. In this or that particular casewhen statistical dispersions are recorded while some set of *stable* global conditions insures definability of a random phenomenon, one just asserts – on the basis of symmetries – that an *a priori* equipartition of the individual probabilities assigned to the involved elementary events is justified and that, in consequence of this, the probability of each event can be defined as the ratio between the number of elementary events by which the considered event realises (the number of favourable elementary events) and the number of all the possible elementary events. This definition, however, cannot be made use of in *any* case, because in general no obvious symmetries do come in and the *a priori* equipartition of the elementary events is *not* confirmed by the *a posteriori* effectively counted relative frequencies of the outcomes of these elementary events. While, as mentioned before, even a clear definition of the 'elementary' character, or not, of a given event, is lacking (Mugur-Schächter, M. [1992C] pp. 305-311).

This problem still keeps quite confidential. For the majority of physicists, for the specialists of communication, for the mathematicians who only make use of the theory of probabilities without placing it in the heart of their research, for the men in the street, a profane confidence reigns that all the important questions concerning probabilities certainly have since a long time obtained an answer in the specialised works. Beliefs of this sort arise concerning any scientific question. They are the fragile but necessary ground on which the evolution of science quietly rolls.

---

[1] An algebra built on a set *S* is a set of subsets of *S – S* itself and the void set *Ø* being always included – which is such that if it contains the subsets *A* and *B*, then it also contains $A \cup B$ and $A - B$.

[2] A probability measure defined on *τ* consists of a set of real numbers *p(A)*, each one associated to an event *A* from *τ*, such that: $0 \leq p(A) \leq 1$, *p(U)=1* (norm*), p(Ø)=0*, and $p(A \cup B) \leq p(A)+p(B)$ where the equality obtains iff *A* and *B* are 'independent' in the sense of probabilities i.e. iff they have no elementary event $e_i$ in common *(A∩B=Ø)*. The number *p(A)* yields the value of the limit – supposed to exist – toward which the relative frequency *n(A)/N* converges when the number *N* of realisations of the involved repeatable procedure *Π* is increased toward infinity, *n(A)* being the number of outcomes of *A* when *Π* is repeated *N* times.

But those who develop a research involving the foundations of the theory of probabilities are entirely conscious that today the abstract concept of a probability measure entails a vitally important problem of interpretation. Kolmogorov ( [1963]) himself wrote (quoted in Segal [2003]) :

> « I have already expressed the view …that the basis for the applicability of the results of the mathematical theory of probability to real random phenomena must depend in some form on the *frequency concept of probability*, the unavoidable nature of which has been established by von Mises in a spirited manner…..(But) The frequency concept (of probability)[3] which has been based on the notion of limiting *frequency* as the number of trials increases to infinity, does not contribute anything to substantiate the applicability of the results of probability theory to real practical problems where we have always to deal with a finite number of trials ».

This quotation deserves much attention. One cannot be clearer. Nevertheless let us comment. At the present time there exists a more or less fuzzy but very active belief according to which the law of big numbers would establish in a *deductive* way the existence, for any factual random phenomenon, of a factual probability law of which, moreover, also the structure would be specified. But *this is false*. The well known theorem of big numbers asserts only what follows (I make use of the traditional notations).

Given a set $\{e_j\}$, $j=1,2,....q$ of events $e_j$ (or of elementary events, indifferently), *if* a factual probability law $\{p(e_j)\}$, $j\equiv 1,2, ....q$ on this set *does exist, then* for every $e_j$ and every pair $(\varepsilon,\delta)$ of two arbitrarily small real numbers, there exists a whole number $N_0$ such that when the number $N$ of 'identical' reproductions of the experiment $\Pi$ from the considered random phenomenon becomes equal to, or bigger than $N_0$, **(**the meta-PROBABILITY

$$\boldsymbol{P}[(\,|n(e_j)/N - p(e_j)\,|) \leq \varepsilon ] \qquad (1)$$

of the meta-event consisting of the fact that (the absolute value of the difference $(n(e_j)/N – p(e_j))$ between, on the one hand the relative frequency $n(e_j)/N$ counted for the event $e_j$, and on the other hand the factual probability $p(e_j)$ assumed for that event, be *smaller* than or equal to $\varepsilon$ )**)** – so the meta-probability of this meta-event – becomes itself *bigger* or equal to $(1-\delta)$. This can also be expressed in a rigorous and more synthetic manner by the following entirely symbolic writing :

$$\forall j, \ \forall (\varepsilon, \delta), \quad (\ \exists N_0 : \ \forall (N \geq N_0)) \ \Rightarrow \ \boldsymbol{P}[(\,|n(e_j)/N – p(e_j)\,|\,) \leq \varepsilon ] \geq (1 - \delta) \quad (2)$$

This same assertion is sometimes expressed less precisely by saying that *if* a probability law $\{p(e_j)\}$, $j=1,2,....q$ does exist on the set of events $\{e_j\}, j=1,2, ...q$, *then* for any $j$, as $N$ 'tends toward infinity', the absolute value of the difference between the relative frequency $n(e_j)/N$ and the probability $p(e_j)$, 'tends in probability' toward *0*, i.e. it tends *nearly* certainly toward *0*. *Nearly* certainly, *not* certainly, because in the expression $\boldsymbol{P}[(\,|n(e_j)/N - p(e_j)\,|) \leq \varepsilon ]$ the symbol $\boldsymbol{P}$ designates itself only a meta-*probability*, not a certainty [4].

So in the theorem of big numbers the existence of a factual probability law is by no means proved, it is posited. What is proved indeed is that if a probability law $\{p(e_j)\}$, $j=1,2....q$ does exist, then, as the number $N$ of the achieved trials increases, the tendency of each relative frequency $n(e_j)/N$ of an event $e_j$ toward the probability $p(e_j)$ assigned to $e_j$ by that supposedly existing law, is itself very 'probable' *in the sense of another factual probability law* designed by the symbol $\boldsymbol{P}$, which is also posited to exist. So concerning the significance of the existence of a factual probability law, the theorem of big numbers offers only an infinite regression.

As for the structure of the posited factual probability law $\{p(e_j)\}$, $j=1,2....q$, the theorem of big numbers constructs indeed a factual definition of it – the famous 'relative frequency definition' expressed by the use of the meta-probability $\boldsymbol{P}$ – but this definition : *(a)* is non effective (as Kolmogorov stresses) ; and *(b)* it is constructed on the basis of the postulation of the existence of the probability law $\{p(e_j)\}$, $j=1,2....q$, without in any way specifying what this 'existence' means, in what physical features of what sort of physical entity it consists. Indeed in *(2)* the relative frequencies $n(e_j)/N$ can be conceived to play a role of 'determination' of the limiting numerical values $p(e_j)$, only if these relative frequencies are *subject* to the postulated existence of the undefined and unexplained limiting values $p(e_j)$.

---

[3] My specification

[4] Throughout these formulations the prefix 'meta' means that the definition of the considered event or probability *involves*, respectively, the events $e_j$ and the probabilities $p(e_j)$ and therefore it is conceptually posterior to these.

This is the conceptual situation toward which point Kolmogorov's above quoted assertions.

And indeed, when one concentrates attention upon this situation it leaps to one's eyes that as long as a clearly and *independently* constructed concept of a *factual* probability law is lacking, it is improper to relate probabilities – in the factual sense of the term – with a formal system like Kolmogorov's mathematical theory of probabilities.

*What is lacking is a model of the concept of factual probability law*. The abstract definition of a probability measure has to be the *formalisation* of such a factual model , it cannot be its *generator*.

Confusions of this kind, between the conditions to be required – specifically – concerning the existence and the performances of descriptions of *factual* entities and, on the other hand, the conditions to be required concerning a purely syntactic system of symbols concerning these descriptions of factual entities, are not rare. And, systematically, such confusions introduce long lasting false problems.

Already before Kolmogorov, other authors also have manifested reservations with respect the applicability of Kolmogorov's theory of probabilities. For instance R. J. Solomonoff [1957] wrote :

« Probability theory tells how to derive a new probability distribution from old probability distributions…….. It does *not* tell how to get a probability distribution from data in the real world ».

But it was Kolmogorov himself who finally developed a definitive veto concerning the applicability of his mathematical theory, to factual problems. Throughout the decade 1980 he expressed refusal of Shannon's central concept of 'informational entropy'[5]. Quite radically Kolmogorov [1983] has advocated *the elimination of his formal concept of probability, from all the representations which had been considered as applications of this concept* :

«**1**. Information theory must precede probability theory and not be based on it. By the very essence of this discipline, the foundations of information theory have a finite combinatorial character.

**2**. The applications of probability theory can be put on a uniform basis. It is always a matter of consequences of hypotheses about the impossibility of reducing in one way or another the complexity of the descriptions of the objects in question. Naturally this approach to the matter does not prevent the development of probability theory as a branch of mathematics being a special case of general measure theory.

**3**. The concepts of information theory as applied to infinite sequences give rise to very interesting investigations, which, without being indispensable as a basis of probability theory, can acquire a certain value in the investigation of the algorithmic side of mathematics as a whole».

So the father of the modern mathematical theory of probabilities wanted the informational problems as well as those concerning complexity, to be treated from now on without making use of the concept of probability. He wanted them to be treated by the means of, exclusively, combinatorial analyses of « hypotheses about the impossibility of reducing in one way or another the complexity of the descriptions of the objects in question » (this concerns the definition of the 'elementarity' of entities or events). As for probabilities, he wanted to confine them inside the purely mathematical general measure-theory. He conceived to imprison in an abstract cage the concept of probability so profoundly rooted into the concrete human experience!!! This is a proposition made by a major thinker, so it has to be seriously taken into account. But it is an extremist proposition.

Among mathematicians this proposition has been accepted without resistance and it has already changed the direction of research concerning 'complexity' : for mathematicians the physical entities are like shadows of the physical ones.

But for a physicist it is simply not conceivable that a formal concept like that of a probability measure – *which stems from factuality* – be unable to point in return toward an explicitly constructible factual significance. Consider that inside fundamental quantum mechanics the descriptions of microstates emerge in a 'primordially' probabilistic form : on a *tabula rasa* of previous conceptualisation, they emerge directly probabilistic. They emerge probabilistic in the very *first* stratum of the conceptualisation of what is denominated 'microstates'. Inside macroscopic physics the probabilities are conceived as a manifestation of our ignorance of 'details' which, in principle, can be known and which the basic *non* probabilistic theories (mechanics, electromagnetism, etc.) are posited to be in principle able to treat so as to obtain certain results. But the classical theories have *failed* to yield an acceptable representation of microstates. So fundamental quantum mechanics has been constructed directly outside the classical physics and even outside the general classical general thinking (even if in prolongation of these, in certain respects). So beneath the probabilistic quantum mechanical descriptions of microstates there does not exist a more basic non probabilistic theory of microstates able in principle to offer assured, certain results about them. Furthermore there are no knowable factual data *exterior* to the probabilistic laws which *directly* and *wholly* constitute the factual content of these descriptions. One might succeed some day to SUPERPOSE to the descriptions of micro-*states* from *fundamental* quantum mechanics, satisfactory deterministic models (as it has been tried indeed since a long time already (without always clearly distinguishing 'states' from 'systems')[6]. But the *conceptual status* which always will have to be assigned to such models, whether successful or not, will be only that of *derived* meta-descriptions founded upon the primary ones built for micro-states inside fundamental quantum mechanics. *They cannot be the primary descriptions* : there occurs here a kind of inversion of the stratification which we used to admit inside classical thinking where one *begins* with already constituted *models* – 'objects' : a 'mobile', a 'gas', a 'charged particle', a 'compound molecule', etc. – introduced as a *primary* datum, and *afterward* constructs descriptions of the *states* of these, mechanical, electromagnetic, thermodynamic states, etc. (The long-lasting absence of a clear success in the domain of elementary particles and micro-gravitation might stem from precisely the fact that this inversion and its specific consequences are not noticed, while on the other hand the semantic contents of fundamental quantum mechanics itself are still far from having been thoroughly understood).

---

[5] The mathematical expression $H(S)=\Sigma_i p_i log(1/p_i)$ which possesses the same form as Boltzmann's function of physical entropy $\mathbf{S}=\Sigma_i(n(e_j)/N)log(1/(n(e_j)/N))$ but where, instead of the relative frequencies $(n(e_j)/N)$ of a set of factual events $\{e_j\}$, $j=1,2, …,…q$, are inserted the probabilities from a probability measure $\{p_i\}$, $i=1,2,…q$ in the sense of Kolmogorov's theory, defined on a set of signs $\{a_i\}$, $i=1,2,…q$ emitted by a 'source of information' in order to be coded and utilised for the transmission of messages.

So for a physicist trained in, specifically, fundamental quantum mechanics, where one deals exclusively with directly probabilistic descriptions of micro-states, it simply seems absurd to conceive that the concept of a factual probability law be not constructible, when this concept is so organically involved, with such a primordial conceptual status, in the very basis of the entire nowadays microphysics, which in its turn yields, in principle at least, the foundation of our whole physical knowledge. For such a physicist the only receivable formulation of the problem entailed by the present conceptual situation concerning probabilities, is : define a model of a factual probabilistic situation, such that it assigns a definite significance to the *existence* of a factual probability law and that it permits to effectively specify the *structure* of this law.

### III. BRIEF PRELIMINARY CONCERNING A
### *METHOD OF RELATIVISED CONCEPTUALISATION (MRC)*

I have shown elsewhere (Mugur-Schächter [1991],[1992B],[2002A],[2002B],[2006]) that Kolmogorov's formalisation of probabilities, though it has remarkably enriched the preceding representations of the concept of probabilities, does *not* subtend the most general concept of probability. For proving this assertion it suffices to produce an example. Now the example that can be evoked is indisputable and huge : the probabilistic descriptions of microstates *exceed* the classical concept of probability ; this classical concept is entirely overflowed by the descriptions constructed in fundamental quantum mechanics. So it is obvious that certain data necessary for fully identifying the *global* contour of the concept of probability and in particular the 'significance' of the central probabilistic concept of a factual probability law, remained hidden to the classical representations of the concept of probability.

But these additional data can be discerned *via* an appropriate more exhaustive analysis. Such an analysis has been effectively developed inside a *general method of relativised conceptualisation (MRC)* (Mugur-Schächter [1984], [1992B], [1992C], [1995], [2002A], [2002B], [2006]). The aim of this method is to build a system of norms of conceptualisation which exclude by construction the possibility of emergence of any false problems or paradoxes. The method constrained by this goal is founded upon a appropriate generalisation of the semantic essence of the representation of microstates involved in the formalism of fundamental quantum mechanics. Indeed, once explicated (Mugur-Schächter [2008]), this semantic essence appears to have incorporated the *universal* features of a very first *omnipresent* stratum of human conceptualisation of which the very existence, so *a fortiori* the structure, had remained entirely non perceived, occulted beneath the classical thinking such as it is manifested by the current languages as well as by classical logic and probabilities (Mugur-Schächter [2006]). In what follows, for self-sufficiency of this work, we introduce a very brief enumeration of exclusively those features of *MRC* that will be made use of in the subsequent construction of a model of the concept of factual probability law. (Detailed presentations can amply be found in the last three works on *MRC* mentioned above as well on my web site (at http://www.mugur-schachter.net/publications.html ).

---

[6] Concerning micro-*systems* (microscopic entities defined by permanent 'properties' assigned to them (mass, charge, spin, etc.) the conceptual situation is different : in the atomic physics they were directly defined as models, and in the physics of elementary particles the same attitude is practiced)

By the very definition of the concept, a 'description' necessarily involves an entity playing the role of *object-of-description* (the object-entity), and a grid for qualifying this object-entity (the qualifying 'view'). The method of relativised conceptualisation – *MRC* – introduces a canonical, a standard, a normed descriptional form where the considered object-entity and the qualifying view are explicitly referred to while, on the other hand, their *structures* are specified by explicit definitions. More precisely :

*(1)* MRC is founded upon a systematic relativisation of any description, to a triad $(G, œ_G, V)$ where $G$ denotes the *operation of generation* – physical, or abstract or consisting of some combination of physical and abstract operational elements – by which the object-entity is made available for being qualified ; $œ_G$ denotes the *object-entity* itself introduced by $G$ ; $V$ denotes the *view* by which the object-entity is qualified.

*(2)* Any description is denoted by the symbol $D/G, œ_G, V/$ which explicitly points toward the non removable relativities to the particular triad $(G, œ_G, V)$ which it brings in[7].

*(3)* A one-to-one relation $G \leftrightarrow œ_G$ is posited between the operation of generation $G$ and the object-entity $œ_G$ introduced by $G$. This is a *methodological posit* that is far from being obvious. But upon analysis it has been found to impose itself inescapably concerning the quantum mechanical descriptions of microstates. And an attentive further examination established that – if one wants to erect a method of conceptualising that banishes *a priori* any insertion of false absolutes – this methodological posit also imposes itself inescapably with *full generality* inside the whole class of first-stratum descriptions of *any* nature. Precisely this inescapable character of a one-to-one relation $G \leftrightarrow œ_G$ entails major conceptual consequences concerning the 'primordial' sort of probabilities brought in by first-stratum descriptions ((Mugur-Schächter [2006] pp.61-66 and 213-221, [2008])).

*(4)* Any view $V$ is endowed by definition with a strictly prescribed structure. A view $V$ is a *finite* set of *aspect-views* $Vg$ where $g$ is an aspect-index ; an aspect-view $Vg$ (in short : an aspect $g$) is *a semantic dimension of qualification* (colour, weight, etc.) able to carry any *finite* set of 'values' $(gk)$ of the aspect $g$ which one wishes to consider (for colour : red, yellow, green, etc.) (the bracket surrounding $gk$ shows that this symbol functions like a unique index). (In a definite case the indexes $g$ and $(gk)$ can be replaced by any other convenient signs). An aspect-view $Vg$ is defined *if and only if* are defined all the devices (instruments, apparatuses) as well as all the material or abstract operations on which is based the assertion that an examination of any given object-entity *via* the aspect-view $Vg$, has yielded this or that – unique and definite – value $(gk)$ (or none). A view $V$ is a finite *filter* for qualification : *with respect to aspects, or values of aspects, that are not contained in it by its initially posited definition, a given view $V$ is blind, it does not perceive them*. The qualifications of space and time are achieved *via* a very particular sort of *frame-views* $V(ET)$ (reducible, if convenient, to a space-frame-view $V(E)$ only or a time-frame-view $V(T)$ only).

*(5)* Given a pair $(G, Vg)$, the two epistemic operators $G$ and $Vg$ can *mutually exist*, or not. If any examination of the object-entity $œ_G$ introduced by the object-entity generator $G$ produces one well defined result $(gk)$, then the aspect $g$ and the aspect-value $(gk)$ of $g$ both do exist with respect to $G$, i.e. there is *mutual existence* between $G$ and $Vg$. In this case $(G, Vg)$ is *an epistemic referential*. This means that in this case, if one applies to the object-entity $œ_G$ introduced by $G$, an examination by $Vg$, so if one produces operational successions $[G.Vg]$, then one obtains the relativised description $D/G, œ_G, Vg/$ of $œ_G$ *via* the grid for qualification consisting of the aspect-view $Vg$.

If on the contrary, what is defined to be an examination by $Vg$, when applied to the object-entity $œ_G$, yields no definite result, then there is no mutual existence between $Vg$ of $œ_G$ ($œ_G$ does not exist relatively to $Vg$ and *vice versa*). In this case the matching $(G, Vg)$ has to be eliminated *a posteriori* as unable to generate a relative description $D/G, œ_G, Vg/$.

If, after some number $N$ of repetitions of the succession $[G.Vg]$[8] only one and the same value $(gk)$ of the aspect $g$ is systematically obtained, the corresponding relative description $D/G, œ_G, Vg/$ is *'N-individual'*, $N$ being always finite (in short, an individual description). If on the contrary the obtained value $(gk)$ in general varies from one realisation of $[G.Vg]$ to another one, the corresponding relative description $D/G, œ_G, Vg/$ is *statistical*, so *via* a very big but finite number $N'$ of $N$ repetitions of $[G.Vg]$ it can *'N'-point'* toward a *probabilistic* description $D/G, œ_G, Vg/$ (cf. Mugur-Schächter [2006] pp. 75-78) : all the involved concepts are kept finite, effective.

These considerations can be extended in an obvious way to also any pair $(G, V)$. In this case one speaks of the possibility, or not, of a relative description $D/G, œ_G, V/$ which, if it does exist, can be individual or statistical-probabilistic.

---

[7] This last relativity cannot be absorbed into that concerning $G$ : the qualifications depend directly on $œ_G$ and cannot be derived from $G$.

[8] In *general*, after a succession $[G.Vg]$ the replica of the object-entity $œ_G$ involved in that succession either is changed by the examination via $Vg$, or it is destroyed (absorbed in a device, etc.). So repetitions of $[G.Vg]$ require repetitions of also G (creation of a new replica of $œ_G$).

*(6) The space-time frame-principle* asserts what follows concerning – specifically – physical object-entities. *Any physical object-entity does exist relatively to at least one aspect-view Vg that is different from any space-time frame-view V(ET) ; but it is in-existent with respect to any space-time frame view V(ET) considered alone, separately from any aspect-view Vg different from any space-time aspect ET*. Consider then a physical object-entity $œ_G$ generated by a physical operation $G$ : in consequence of the space-time frame-principle the view $V$ from any epistemic referential $(G,V)$ able to generate a description of $œ_G$ must include a space-time frame-aspect $V(ET)$ as well as at least one aspect-view $Vg$ different from any space-time aspect $ET$ yielding a partial relative description $D/G,œ_G,Vg/$ of $œ_G$.

*(7)* The points *(5)* and *(6)* entail what follows. Since a view $V$ is a union of a finite number $m$ of aspect-views $Vg$ we can write $V=\cup_g Vg$, $g=1,2...m$. Each aspect-view $Vg$ introduces a *semantic g-axis* that carries its 'values' *(gk)*, $k=1,2,...w(g)$ where $w(g)$ is an integer that depends on $g$. So $V$ introduces the abstract *representation space* defined by these $m$ semantic $g$-axes and *any relative description $D/G,œ_G,V/$ consists of a cloudy structure or 'form' of (gk)-value-points with $g=1,2...m$, $k=1,2,...w(g)$ contained in the m-dimensional representation-space of the view V which it introduces*. If the object-entity $œ_G$ is of a physical nature, one must add inside $V$ a *4-dimensional discreet space-time view V(ET)* and the relative description $D/G,œ_G,V/$ becomes a cloudy structure or 'form' of *space-time-(gk)-value-points* with $g=1,2...m$, $k=1,2,...w(g)$, and $x,y,z,t$, some finite space-time grid upon which the units of space and time impose a discrete set of possible space-time values, this whole form being contained in the *(m+4)-dimensional representation-space* now introduced by of the view $V$.

*(8)* One can form *descriptional chains* i.e. chains of descriptions connected *via* common elements in their object-entities or in their views. Along a descriptional chain there exists a *descriptional hierarchy* : the order *1* is assigned to the first description from that chain ; the second description connected to the first one is of order *2* with respect to it (a *meta-description*[9] with respect to the first one) ; the third description is assigned the order *3* and it is a meta-description with respect to the description of order *2* and a *meta-meta*-description with respect to the first description from the chain). Etc. So in general the order of a description inside a chain is relative to the process of construction of the chain. But if the considered chain starts with a basic first-stratum description, then this description marks an *absolute* beginning of a process of construction of knowledge and the order *0* is assigned to it.

*(9)* Passage from a given description belonging to a descriptional chain, to the following one, is commanded by *the principle of separation*, in the following sense. Each relative description $D/G,œ_G,V/$ is accomplished inside an epistemic referential $(G,V)$ where $G$ – in consequence of the methodologically posited one-to-one relation $G\leftrightarrow œ_G$ – is tied to *one* object-entity $œ_G$ and the view $V$ consists of a *finite* set of aspect-views $Vg$ each one of which carries a *finite* set of aspect-values *(gk)*. Furthermore the relative description $D/G,œ_G,V/$ is achieved *via* some finite number of realisations of successions *[G.Vg]*. So a relative description $D/G,œ_G,V/$ is by construction a *finite* 'cell of conceptualisation' : after the realisation of some finite number of successions *[G.Vg]*, the descriptional resources from the epistemic referential $(G,V)$ are *exhausted*. Then if one wants to obtain some knowledge not obtained by $D/G,œ_G,V/$, one has to somehow bring in *another* epistemic referential $(G,V)'$ different from $(G,V)$ and to construct the *new* relative description $(D/G,œ_G,V/)'$ corresponding to $(G,V)'$. Now the *principle of separation requires that $(D/G,œ_G,V/)'$ be always achieved by a process explicitly and entirely separated from the descriptional process which led to $D/G,œ_G,V/$*, thereby systematically avoiding any coalescence or confusion between the geneses of two distinct relative descriptions.

*(10)* According to *MRC* any knowledge that can be communicated in a *non* restricted way, is *description* ('pointing toward' restricts to real or virtual co-presence inside some delimited space-time domain, so are also mimics, emotional sounds, etc.). Nothing else but descriptions can be unrestrictedly communicable knowledge, neither 'facts' which are exterior to any psyche, nor psychic facts (emotions, desires, etc.) which are not expressed by some more or less explicit description, verbal or of some other constitution. In particular :

> When the concept of probability is reconstructed inside *MRC*, the elementary events and the events from any probability space acquire the conceptual status of relative *DESCRIPTIONS* : their *MRC*-status is *not* that of object-entities $œ_G$, it is that of *relative DESCRIPTIONS of object-entities $œ_G$*.

This will reveal high and non trivial powers of organisation in what follows.

This very condensed sequence of extracts from *MRC* will suffice for sketching now out how the descriptional relativisations required by *MRC* permit to associate a model to the concept of a factual probability law. So what follows can be regarded as an illustration of the way in which the method of relativised conceptualisation works. Those who will desire to place the subsequent development inside a fully elaborated *MRC* context, are referred to (Mugur-Schächter [2006] pp.197-257).

---

[9] In logic the verbal particle 'meta' indicates an imbedding language, so it is conceived as placed 'under' the studied language. Here, on the contrary, 'meta' is assigned the significance of 'after'-and-connected-with.

## IV. CONSTRUCTIVE OUTSKETCH OF
## A MODEL OF THE CONCEPT OF A FACTUAL PROBABILITY LAW

### IV.1. Preliminary : games with a parcelled painting

This preliminary investigation will consist of a succession of examples. By passage from a small obviousness to another small obviousness there finally will emerge a novelty : a definition of the factual probability law to be asserted in a particular case of probabilistic situation.

#### *IV.1.1. Relativised parcelling and notations*

Consider the puzzle of a painting *P* representing a landscape, *100* square pieces $\sigma$. Consider also a spatial grid that can be superposed to the integrated solution of the puzzle of *P*. On this grid each square $\sigma$ is localised by the specification of its two space coordinates $(x_k, y_h)$ where $x_k$ is an element from a set of *10* successive equidistant coordinates $\{x_k\}$, $k=1,2...10$ marked on a horizontal space axis *ox* superposed to the lower edge of the painting *P*, while $y_h$ is an element from a set of *10* successive equidistant coordinates $\{y_h\}$, $h=1,2...10$ marked on a vertical space axis *oy* superposed to the vertical left hand side edge of *P*. The label $(x_1, y_1)$ indicates the square from the left lower corner of *P* and the left lower corner of this square is the origin *0* of the plane Cartesian system of reference axes *xoy* attached to the grid superposed to *P* ; while the pair $(x_{10}, y_{10})$ indicates the square from the right upper corner of *P*.

Consider an epistemic referential $(G_P, V)$ where the object-entity generator $G_P$ is a 'selector' that selects as an object-entity the integrated solution of the puzzle of the painting *P* and *V* is a view which consists of three aspect-views defined as follows :

* The space-frame-view is the union $V(E) \equiv V(El) \cup V(E\phi)$ where : $V(El)$ is, specifically, a frame-view of *spatial location* of which the possible values are the *100* pairs of spatial coordinates $(x_k, y_h)$, $k=1,2,...10$, $h=1,2,...10$ (so a square $\sigma$ examined *via* the view $V(El)$ leads to the description $D/G_\sigma, \sigma, V(El)/$ of spatial location of $\sigma$ consisting of one among the pairs of coordinates $(x_k, y_h)$, $k=1,2,...10$, $h=1,2,...10$) ; $V(E\phi)$ is a frame-view of *spatial form* endowed with a very big number of 'values of form' (this amounts to the introduction of a very small unit of length that permits to reproduce satisfactorily any perceivable contour).

* A colour aspect-view $Vc$ endowed with a set of colour-values rich enough for insuring that a relative description $D/G_\sigma, \sigma, Vc \cup V(E\phi)/$ yields a form-of-colour covering the object-entity $\sigma$ which reproduces 'satisfactorily' that one perceived on $\sigma$ by a normal human eye. (However, since everything in the definition of any view is by construction discrete and finite while any view acts like a filter, the total number of possible distinct 'values-of-colour-form' is discrete and finite). The view $Vc \cup V(E\phi)$ can be synthetically rewritten as $Vc \cup V(E\phi) \equiv Vc\phi$ where $V\phi$ is a *view of colour-form*.

With these definitions and notations the description of the integrated puzzle of the painting *P* achieved inside the epistemic referential $(G_P, V)$, has to be written as as $D/G_P, P, V(El) \cup Vc\phi /$.

Consider now a 'local' epistemic referential $(G_\sigma, V)$ where $G_\sigma$ selects as object-entity only *one* square $\sigma$ while the view *V* is the same one is in the referential $(G_P, V)$. Then a relative description corresponding to $(G_\sigma, V)$ is to be written as $D/G_\sigma, \sigma, V(El) \cup Vc\phi/$ : it consists of a 'colour-form' covering the selected square $\sigma$ and which is located as indicated by its 'value' $(x_k, y_h)$ of spatial location.

Let *Vac* be a new *'approximate-colour'* view endowed with *q uniform* approximate-colour values *j*, $j=1,2,...q$ (this square is approximately of this uniform shed of red, that square is approximately of this uniform shed of blue, etc.).

If in the local relative description $D/G_\sigma, \sigma, V(El) \cup Vc\phi/$ of a square $\sigma$, the view $V(El)$ of spatial location is cancelled, one obtains a local description $D/G_\sigma, \sigma, Vc\phi/$ of a square $\sigma$ where any *direct* indication of spatial location is filtered out.

If furthermore, in this new local relative description $D/G_\sigma, \sigma, Vc\phi/$, the view of colour-form $Vc\phi$ is replaced by the view *Vac* of uniform approximate-colour, the value of colour-*form* that covered the considered square $\sigma$ is equally filtered out and a new relative description arises – to be written as $D/G_\sigma, \sigma, Vac/$ – where, in consequence of the *uniformity* of the approximate-colour values assigned to *Vac*, one looses now also the perceptibility of any affinity or repulsion between the form-of-colour reaching a border of the considered square $\sigma$, with respect to another form-of-colour reaching another border of another square $\sigma$. So one ceases to be able to play puzzle with the *100* squares described *via* exclusively *Vac* : this time any hint of some connection between the considered square $\sigma$ and the global 'significance' carried by the integrated painting *P*, is lost.

Suppose now that the global dimensions of the picture $P$ and the distance between two successive values of the $x_k$ or $y_h$ coordinates, are *such* that :

*(a)* Any square $\sigma$ is small enough for carrying only *one* approximate-colour-value $j$. Then its relative description $D/G_\sigma,\sigma,Vac/$ *via* the view $Vac$ of uniform approximate-colour entirely consists of only one uniform approximate-colour : it reduces to just its unique approximate-colour value $j$. So we can write $D/G_\sigma,\sigma,Vac/\equiv Dj$, $j=1,2,....q$. Then we have $\{D/G_\sigma,\sigma,Vac/\}\equiv\{Dj\}\equiv\{j\}$, $j=1,2,....q$.

*(b)* Any given partial description $Dj \equiv j$ is realised on *much* more that only one square from $P$. Thereby, by construction, the cardinal $q$ of the set of *mutually different* relative descriptions $\{Dj\}\equiv\{j\}$, $j=1,2,....q$ is *much smaller than 100*.

So finally, each one among the *100* squares $\sigma$ can be taken knowledge of *via* three distinct views : the frame-view $V(El)$ of spatial location , the frame-view view of $Vc\phi$ of colour-form, and the view $Vac$ of uniform approximate-colour.

Let us now mix the squares and throw them all into a ballot box.

Starting from this point we define a succession of 'games' which will lead to the announced interesting conclusion.

### *IV.1.2. Game illustrating the power of reconstruction contained in space (or space-time) order*

Let us accomplish the *100* possible successive extractions of a square $\sigma$ from the ballot box and look at each extracted square *via* the frame-view $V(El)$ of spatial location. This, for each square, yields a description which places each square at the place, on the reference grid subtended by axes *xoy*, which is indicated by the obtained coordinates $(x_k,y_h)$. Since any view acts is a filter, this happens *without having taken into account the colour-form carried by it, nor the uniform approximate-colour j defined on it by the view Vac*. Nevertheless, after exactly *100* extractions, the global painting $P$ is reconstructed. Though the order of extraction of the squares will have been random, each individual act of progression toward the reconstruction of the global painting $P$ will have been accomplished in a way marked by *certainty*, while the global process will have been *finite* : the spatial grid of reference possesses a power of topological organisation which is *independent of the 'semantic content' of the squares*.

These remarks extend in an obvious way to the case also of a space-time grid.

### *IV.1.3. Puzzle with only one replica of $P$*

Let us now proceed differently. Let us again mix the squares and shed them into the ballot box. Let us then make again the *100* possible successive extractions of a square from the ballot box. But this time, let us make use of – exclusively – the view $Vc\phi$ of colour-form. So each square is perceive by its relative description $D/G_\sigma, \sigma, Vc\phi/$. The *label of space location* $(x_k,y_h)$ as well as well as its label $j$ of uniform approximate-colour are filtered out, they are ignored. In these conditions, again, after exactly the $100^{th}$ extraction the global painting $P$ will be reconstructed. But in general, for finding the right place where to put an extracted square, we will have had to fumble around by trials and errors ; but, guided by the structure of the form-of-colour carried by the square, we will have finally identified the 'good' place of the square. And the structure of the form-of-colour of the square will have been useful mainly by its content in the proximity of the *borders* of the square where, for each given border, it determines a sort of *neighbourhood-coherence* with the form-of-colour reaching a *unique* other border of another square. A sort of attraction by semantic continuity acts between the two mentioned borders and, on the contrary, a sort of repulsion by semantic discontinuity works between the form-of-colour that reaches the initially considered border of the considered square, and any other form-of-colour reaching any border of any other square. This time the independent power of topological organisation of the space coordinates will have been filtered out and replaced by these 'attractions *via* semantic continuity' or 'repulsions *via* semantic discontinuity'. And again nothing infinite will have been involved and nothing will have been random – if abstraction is made of the randomness in the order of extraction of the squares – notwithstanding the presence of trials and errors. For, quite obviously, the trials and errors are tied with features of the defined situation, of which the nature is radically different from that of a predictive uncertainty in the probabilistic sense.

This example, like the preceding one, can be extended in an obvious way to the case of an '*evolving* picture' fragmented in space-*time* cubes of which the space-time labels are by-passed, while exclusively other descriptional contents are considered, with the attractions by continuity on the borders and the repulsions by discontinuity on the borders which these contents entail. (During the research of a criminal, for instance, in essence, one plays a generalised space-time puzzle game).

### *IV.1.4. Puzzle with several replicas of the painting P*

Let us now provide ourselves with *1000* replicas of the fragmented painting *P* and let us proceed with these in the same way as we did above for one replica : we mix together all the *100 000* squares which we now possess, we shed them all into the pool-box, and then we extract the squares one by one, ignoring the space label and the approximate-colour label imprinted on it and searching for the square an appropriate place on one or other among the *1000* void space-time grids placed in front of us. What will happen ? After *100 000* extractions from the ballot-box we *certainly* shall have entirely reconstructed all the *1000* replicas of the fragmented painting *P*. But this will have been achieved only after quite a lot of trials and errors and not by a neatly separate completion of the replicas, in succession, but by an intermingled process of completion of all the replicas, involving leaps from one replica to another one. In general, only by the last extractions will all the *1000* replicas have entirely separated from one another.

In principle, no essentially new features are brought in by the use, instead of *1000* replicas, of $10^n$ replicas with *n* an arbitrary whole number. And this game also can obviously be extended to a set of 'evolving paintings'. And again nothing infinite will have been involved and – notwithstanding the presence of trials and errors – nothing will have been random if abstraction is made of the randomness in the order of extraction of the squares.

A puzzle game, no matter how complex and big, involves randomness exclusively in the order of extraction of the squares.

*The attractions by semantic continuity on the borders of the squares and the repulsions by violation of semantic continuity on the borders, exclude randomness from the final reconstruction of any number of replicas of the global entity which has been parcelled.*

### *IV.1.5. A probability game with one replica of the painting P*

How, then, does 'probabilistic randomness' emerge ? By a modification which, at a first sight, will seem insignificant, suddenly all the characters of a 'probabilistic situation' will come in : unending sequences of elementary events, the corresponding statistical relative frequencies, probabilistic randomness and probabilistic convergence. The announced apparently insignificant modification will reveal itself to have been in fact a radical conceptual jump.

We make use of the same parcelled painting *P* involved in the preceding paragraphs. But this time, instead of a puzzle game, let us play, with just one replica of this puzzle, the following 'probabilistic game'. Let us mix the squares and shed them into the ballot box. Then let us extract a square. Let us use exclusively the view *Vac* of uniform approximate colour, note the value of the index *j* that appears in the corresponding relative description $D_j, j=1,2....q$, and then *drop the examined square back into the ballot box* (both the aspect *Vϕc* of colour-form and the space frame-aspect *V(El)* of spatial location remain dumb, *so a fortiori* the semantic continuity on the borders of the square remain inactive). Let us then mix the squares from the ballot box and repeat the same procedure an arbitrarily big number of times.

I assert that this time, in consequence of the specified modification of the procedure, we find ourselves in a standard 'probabilistic situation'. Indeed, contrary to what happened in all the preceding cases, this time, *before* each extraction, a certain set of *invariant* conditions is reconstituted, which defines – in the usual sense of probabilistic language – a 'reproducible procedure' *Π* and a *stable* universe of elementary events $U \equiv \{j\}$, $j=1,2....q$, so a random phenomenon *(Π,U)*. Since according to *MRC* (point *(9)*) any communicable knowledge is description, let us rewrite explicitly the universe *U* as a universe of relative descriptions $U \equiv \{D_j\}, j=1,2....q$.

What will happen in these new probabilistic conditions ? *Can that be predicted ?*

If the number of successive extractions and droppings back into the ballot box is very much bigger than the number *q* of elements from *U*, one can make the two following rather obvious remarks *R1* and *R2*.

*R1*. Since the whole initial content of the ballot box is reconstituted after each drawing, all the descriptional values $j=1,2,....q$ that had been possible before some given extraction, are equally possible for the following one. From one extraction to another one, no possibility is irreversibly 'consumed', as it happened in the cases considered in the preceding paragraphs.

*R2*. Correlatively, the content of the ballot box is never exhausted. Nothing brings any more to an end the sequence of results which can be obtained by repetition of extraction-dropping. This sequence is of arbitrary length, it can increase 'toward infinity'.

I add now two other assertions which are not perceived as certainties.

The first one is the answer *A1* to the following question *Q1* : "If one continues indefinitely the repetitions of extraction-dropping, will all the *q* values of the index *j* of medium colour show up, or not ?"

The second assertion *A2* is the answer to the following question *Q2* : "If the repetitions of extraction-dropping are continued indefinitely, how will evolve the relative frequency *n(j)/N* of the outcomes of a given value *j* of the index of medium colour ?".

I now assert a *psychological fact* : after a short reflection, the following answers *A1* and *A2* to the questions *Q1* and *Q2* will gain quasi unanimous consensus among persons skilled in current probabilistic thinking.

*A1*. It is *nearly certain* that if the number *N* of repetitions of extracting-dropping becomes big enough, all the *q* values of the index *j* of medium colour will show up.

*A2*. If the number *N* of repetitions of extracting-dropping is increased without *a priori* limitation, then – earlier or later but *nearly certainly* and *for any j* – the relative frequency *n(j)/N* of the outcomes of a given value *j* of the index of uniform approximate-colour will manifest a certain convergence. Namely, the value of *the relative frequency n(j)/N will tend to reproduce the value of the ratio $n_p(j)/100$ which refers the number $n_p(j)$ of squares from the integrated puzzle of the painting P which carry the considered value j of approximate-colour, to the total number 100 of all the squares from the puzzle of P.*

But *why* should there be a convergence ? And quite especially, *why precisely toward this ratio $n_p(j)/100$ defined on P* ? And why, in both formulations *A1* and *A2*, should one assert a 'nearly certainty' instead of, clearly, a certainty ?

Because, more or less explicitly, in the minds of those who adhere to the answers *A1* and *A2*, some equivalent of the following reasoning takes place.

"Since after every extraction and registering of the obtained *j*-value, the square is released back into the ballot box, and since the extractions are allowed to be repeated indefinitely, there exists no basis for strictly excluding *a priori*, concerning a sequence of arbitrarily big length *N*, the outcome of any one among the different possibilities *{j≡1,2,...q}* ; nor, moreover, the outcome of any order of succession of *j*-values from *{j=1,2,...q}* ; nor the outcome of any one among all the global statistical distributions *{n(j)/N, j=1,2,...q}*, $\Sigma_j n(j)/N=1$, of relative frequencies *n(j)/N* that are constructible for a given *N*, with *j*-values belonging to the universe *{Dj}*, *j≡1,2,...q* of relative descriptions. In the conditions of indefinite repeatability that have been posited here, any outcome of any feature that cannot be *a priori* excluded on the basis of some specified reason, has to be *a priori* admitted as possible. These two formulations have the same significance, so any distinction between them would amount to a contradiction. For instance, nothing permits to strictly exclude *a priori* the maximally unbalanced statistical distribution which, for any given *N* and *j'*, is characterised by *n(j')/N=1*, i.e. *n(j') =N, n(j) =0* for any *j≠j'* (with *j'=2* : *2222222222..... N* times). Indeed if in the first extraction it has been possible to find a square carrying *j=2*, since that square has been released back in the ballot box before the second extraction, the same possibility holds also for the second extraction, and so on, indefinitely. But nothing excludes either to find *j≠2*. This entails the answer *A1*.

However we know that the number of squares in the pool box and the number of possible approximate-colour values *j* are both finite and that *any square comes from the puzzle of the integrated painting P*. In these conditions, *before* each extraction it is natural to expect more to find on the extracted square a *j*-value of approximate-colour which, on the integrated painting P, is repeated, say, on *10* different squares, rather than to find the *j*-value of approximate-colour which on the integrated P is repeated, say, on only *2* different squares. What is effectively found *after* an extraction leaves invariant the reasonableness of the specified expectation before that extraction. We must avoid confusion between *a priori* and *a posteriori* as well as between 'possible' and 'probable'. So, since we know that before each extraction the pool box contains *exclusively* a the *100* squares which compose the puzzle of one replica of the painting P, it is natural to expect *a priori* that in a sufficiently long sequence of *j*-results each possible *j*-value be obtained a number of times approximately proportional to the number of squares on which this j-value is realised *on the integrated painting P* ; and to also expect that while the number *N* of accomplished extractions increases, the relative frequency *n(j)/N* be found to tend to converge, for each given *j*-value, *toward the ratio $n_p(j)/100$ realised for that j-value on one integrated replica of the painting P*. In the posited conditions, any different assumption would be devoid of support, while this one – in a certain sense – simply *follows*.

Indeed the global form of one replica of *P* is contained in there, the pool box, even if it is parcelled. So, in the long term, it *must* manifest itself *via* any view that is not entirely blind with respect to it. Now in the conditions of our probability game, the unique active view is the approximate-colour view *Vac* endowed with the possible values *j≡1,2,...q*. This view is not entirely blind with respect to the form of *P*. And in the conditions of our probabilistic game, the *unique* possible manifestation of the global colour-form aspect of *P* that is possible *via* the approximate-colour view *Va*, consists of a set of relative frequencies *{n(j)/N}, j=1,2,...q* which reproduces the set of ratios $\{n_P(j)/100\}$, $j=1,2,...q$ from the global colour-form aspect of $P$. So such a set of relative frequencies is what has to be expected, by default. Which amounts to the asserted convergence.

However this has to be expected nearly with certainty, not with certainty. This is entailed by the conditions which ourselves have posited : these conditions simply *exclude* the assertion that each relative frequency $n(j)/N\}$ will certainly converge toward the corresponding ratio $n_P(j)/100$, so all the more that it will strictly reproduce this ratio.

Indeed it has already been pointed out that any sequence of $N$ results $j$ is possible, even a sequence *kkkkkkkkk.....* of $N$ results $j=k$. But we are reasoning inside the abstract framework of the concept of probability[10], so for the probabilities on a universe of events of any sort, there is a condition of *norm* : the sum of all the probabilities assigned to the events from the considered universe, whatever these be, must be equal to *1*. Consider than a sequence $\sigma_\omega(N,j)$ of $N$ results $j$ where '$\omega$' is an index of *statistical structure* $\{n(j)/N\}$, $j=1,2,...q$ and $N$ is any whole number. For any give $N$ there exists a corresponding *finite* set $\{\sigma_\omega(N,j)\}$, $j=1,2,...q$, $\omega=1,2,......\nu$, of $\nu$ mutually distinct statistical structures constructible with $N$. These constitute a universe of events (meta-events, with respect to the events from $\{j\}$, $j=1,2,...q$). And the probabilities $p(\sigma_\omega(N,j))$, $\omega=1,2,......\nu$ assigned to these new (meta)events are also subject to the condition $\Sigma_\omega\, p(\sigma_\omega(N,j))=1$. So any sequence $\sigma_\omega(N,j)$, while on the one hand it is possible *a priori*, on the other hand it 'consumes' inside the condition $\Sigma_\omega\, p(\sigma_\omega(N,j))=1$ a certain 'quantity of probability'. This interdicts to assign *a priori* certainty (probability *1*) to any given sequence $\sigma_\omega(N,j)$ : if one did this, thereby, contrary to the initial assumption of *a priori* possibility of *any* sequence, he would *a priori* exclude – for the considered $N$, but whatever it be – the possibility of all the sequences $(\sigma_\omega(N,j))$ but only one among them. Which would be a contradiction. So a certain and strict convergence toward *all* the ratios $n_P(j)/100$, is excluded by the very rules of our probabilistic game.

But *nothing*, in the rules of the probability game, does interdict the intuitive notion that with sufficiently big numbers $N$ *each* relative frequency $n(j)/N$ would – nearly certainly – come arbitrarily near to the corresponding ratio $n_P(j)/100$. Which is precisely the answer *A2* to the question *Q2*.

So the quasi intuitive motivations which underly the answers *A1* and *A2* to, respectively, the questions *Q1* and *Q2*, are now explicit.

### IV.2. Explicit definition of a factual probability law in the case of the 'probability game' with the painting $P$

At a first sight, the motivation brought forth above for the answers to the questions *Q1* and *Q2*, seems trivial. But in fact it discloses a conclusion which, itself, is far from being trivial. Indeed from *A1* and *A2* there finally emerges – for the particular case of a probability game with the picture $P$ – an *effective* definition founded upon 'real facts', of the so elusive concept of a factual probability law. It is by reference to the theorem of big numbers that this definition imposes itself. Indeed when one writes

$$\forall j,\ \forall(\varepsilon,\delta),\quad \exists N_0:\quad \forall(N\geq N_0)\ \Rightarrow\ \mathbf{P}[(\,|n(e_j)/N - p(e_j)\,|) \leq \varepsilon\,]\quad (2)$$

a mere identification of terms permits to clearly perceive that this expression of the theorem can be regarded as a rigorous mathematical translation of precisely the partially intuitive and partially 'reasoned' answers *A1* and *A2* : if one sets $e_j\equiv Dj$ and $\{p(Dj)\}\equiv\{n_P(j)/100\}$, $j=1,2...q$, than one obtains the form *(2)*. Indeed the numbers $\{n_P(j)/100)\}$, $j=1,2...q$, satisfy all the conditions currently imposed upon a factual probability law (they are real positive numbers – here *rational* numbers – they obey the norm condition $\Sigma_j\, n_P(j)/100)=1$, etc.).

So, in the case of the probability game with the picture $P$ the set of ratios $\{n_P(j)/100)\}$, $j=1,2...q$ defines on the set of events $\{Dj\}$, $j\equiv1,2...q$ a quite definite and effective factual probability law

$$\{p(Dj)\}\equiv\{n_P(j)/100\},\quad j=1,2,....q \qquad (3)$$

In this particular case, the problem of the construction of a factual probability law constructed "on the basis of real physical facts", has found a solution : Starting from the painting $P$, by a 'probabilisation' involving a puzzle founded on $P$ and a 'simplification' of the descriptions of the pieces of this puzzle by passage from the initial elements $D/G_\sigma,\sigma(x_k,y_h),Vc\cup V\phi\cup V(E)/$ of this puzzle, to the universe of events $\{Dj\}\equiv\{j\}$, $j=1,2,....q$, we have finally constructed a factual probability space in the sense of Kolmogorov

$$[\{Dj\},\ \tau_X,\ \{n_P(j)/100\}],\quad j=1,2,....q \qquad (4)$$

---

[10] It is **not** circular to introduce here such considerations : here only the concept of a *factual* probability law is acknowledged to be still undefined, but the abstract probabilistic syntax introduced by Kolmogorov is accepted, at least as an initial basis.

where $\tau_X$ denotes *any* algebra on *{Dj}*, *j≡1,2...q*, while the factual probability law *{n$_p$(j)/100}*, *j=1,2,....q* is defined directly on the basic universe *{Dj}*, *j≡1,2...q* (which then determines also the factual probability law on $\tau_X$, whatever its specification).

This conclusion, together with the questions *Q1*, *Q2* and the answers *A1*, *A2* which led to it, involve a definite solution to the question, also, of the *significance* to be assigned in this case to the assertion of merely the *existence* of a factual probability law. Indeed in the answers *A1* and *A2*, the belief in the existence of a factual probability law has been founded upon the fact that before each extraction of a square, a whole replica of the parcelled painting *P* was contained in the pool box, and nothing else ; this presence being constrained by the rules of the game to manifest itself to our knowledge only progressively and in cryptic terms, namely *via* the evolving relative frequencies *{n(j)/N}*, *j=1,2,...q* of outcomes of this or that sign *j*, inside sequences of *N* such signs. (I make use of the word 'signs' because, by construction, *j≡Dj* and in *Dj* any trace of *form*-of-colour *D/G,κ($x_k,y_h$),Vc∪V(E)∪Vϕc/*, *k=1,2,...10, h=1,2...10,* carried by a square, has been filtered out by the uniform 'values' of the approximate-colour aspect-view *Vac*, so that any hint of participation in a more integrated structure endowed with a global 'significance' that would exceed that description *Dj≡j*, has become non perceptible, and so in *(4)* any *Dj≡j* acts as only a sign from a set of signs).

We summarise : in the considered case, the colour-form carried by the global painting *P*, together with the way of parcelling *P* and the view defined on the fragments, determine in effective terms both the significance of the assertion of existence of a factual probability law acting on the universe of events *{Dj}*, *j≡1,2...q*, as well as the structure of this factual probability law. In this case we are in possession of a factual model for the concept of a factual probability law. So in this case the abstract concept of a probability measure is not devoid of interpretation.

### IV.3. Generalisation

There naturally arises now another major question : can this result be generalised ? In what follows we sketch out a positive answer.

#### IV.3.1. Introductory remarks

In order to reach the above interpretation in the particular case considered above, it has been necessary to combine elements from two different but related levels of conceptualisation. On the one hand we have dealt with the level on which are confined the observable data belonging exclusively to – strictly speaking – the probabilistic situation from the probabilistic game with the painting *P*. And on the other hand we have dealt with the level where the global colour-form of the painting *P* itself manifests itself. In that case, by construction, these two levels were both directly observable and we have been free to circulate from one to the other and to optimise explicitly the ways of connecting them so as to generate new understanding.

But this was a very particular and artificial circumstance. In general, an entity which is perceived is not perceived as a puzzle composed of fragments with which it is possible to play a probabilistic game ; it is perceived as just this or that whole entity. And when we perceive a probabilistic situation, we do not conceive some global form related with it. Currently our perception and attention are tied to only one level of perception at a time.

However the example of the probabilistic game with the painting *P* suggests that, quite generally, any probabilistic situation might be connectable with some corresponding global form hidden from direct perceptibility but of which the structure determines the factual probability law that acts on the observed universe of events ; and conversely. Let us try to follow this suggestion.

#### IV.3.2. Probabilisation

We begin with the easiest part, namely extraction of probability spaces from any given entity perceived as a whole.

According to *MRC* « any knowledge that can be communicated in a *non* restricted way, is *description* ('pointing toward' restricts to real or virtual co-presence inside some delimited space-time domain, so are also mimics, emotional sounds, etc.). Nothing else but descriptions can be unrestrictedly communicable knowledge, neither 'facts' which are exterior to any psyche, nor psychic facts (emotions, desires, etc.) which are not expressed by some more or les explicit description, verbal or of some other constitution » (point *(10)*).

Furthermore, « any relative description *D/G,œ$_G$,V/* consists of a cloudy structure or 'form' of *(gk)*-values-points, *g=1,2...m, k=1,2...w(g)* contained in the *m*-dimensional representation-space of the view *V* which it introduces. If the object-entity *œ$_G$* is of a physical nature, one must add to *V* a space-time view *V(ET)* ».

Obviously, any such 'form' can be subject to various *relative* 'probabilisations', each one realised in ways quite similar to that which led from the painting $P$, to the probability space *[{Dj},$\tau_X$,{$n_p$(j)/100}],    j=1,2,....q* from *(4)* : the procedure of probabilisation *does* admit of unrestricted generalisation.

### *IV.3.3. Deprobabilisation*

On the contrary, the following approach is not obvious. So, for the sake of clarity, we split it in successive stages.

***Preliminary outline***. Let us start by considering a probabilistic situation, so a given *stable* random phenomenon *(Π,U)* where the procedure *Π*, though it is said to be 'identically reproducible', generates by repetition a whole universe of mutually distinct elementary events. According to *MRC* the identically reproducible procedure (or experiment) *Π* consists of a sequence *[G.V]* where the operation *G* introduces one replica of the object-entity $œ_G$ to be qualified and *V* is an *active* aspect-view (like in a physical measurement process) which *creates* deliberately the whole action of examination of the replica of $œ_G$ introduced by *G*, thereby leading to a qualification of this replica (Mugur-Schächter [2006] pp. 193-202) by a value of an aspect-view from *V*. So – besides the replica of the object-entity $œ_G$ introduced by the considered realisation of *G* – each realisation of the experiment *Π≡[G.V]* involves various material objects or devices, as well as the conceptual or physical operational elements to be made use of for an examination *via* one or the other among the aspect-views *Vg* that constitute the view *V*. Any such qualification, imprinted by *V* upon the *final* observable state of the considered replica of the object-entity $œ_G$ [11], contributes to the structure of one among the possible elementary events from the universe *U* introduced by the random phenomenon. Now according to *MRC* each entity that can be known in a communicable consensual way, *is a relative description*. But in the case of a random phenomenon *(Π,U)* it is a *simplified* relative description consisting of only a value of some 'label-aspect' drawn *a posteriori* from the effectively perceived description of the elementary event produced by a realisation of *Π≡[G.V]* (in the case of the probability game with the painting $P$, the universe *U≡{Dj}≡{j},   j=1,2....q* has been constructed indeed *via* the simplifying approximate-colour aspect-view *Vac* that filtered out any other aspect beside one 'label-value' *j* from the set of *q* label-values assigned to the simplified aspect of approximate uniform colour).

*This is a crucial feature of the universe U from the definition of any random phenomenon (Π,U).*

Let us then introduce, for any random phenomenon *(Π,U)*, the generalised notation *U≡{Dr}, r=1,2,....s* and let us now focus on the consequences of the feature toward which points the term 'simplified' employed above. When a probabilistic situation is 'given' as a *primary* datum, one does not know the factual probability law that works on the universe *U* of elementary events. Precisely this is the problem examined here. *A fortiori* one has not the slightest awareness of some 'global form' *Φ* permitting to derive from it this factual probability law in a way similar to that made use of in the probability game founded on painting $P$. In a certain sense it would even be misleading to think that such a form always 'pre-exists' in the same way as in the case of the painting $P$.

Nevertheless the bare assertion that one considers oneself to be in a 'probabilistic situation' entails already that one *posits* the *existence* of a factual probability law. To this posited existence of an unknown factual probability law we want to associate a definite and effective definition derived from a convenient description of an integrated form *Φ* globally connected with the considered stable random phenomenon *(Π,U)*. We want to *geometrise* the probabilistic situation represented by the random phenomenon *(Π,U)*, to eliminate from it, by a convenient process of integration, the parcelled, successive, temporal features ; but to eliminate these features in a way *such* as to conserve, on the integrated individual description of the global form *Φ*, a network of constitutive parcelled relative descriptions that shall permit to specify from it, by mere *counting*, the factual probability law to be asserted on the universe of (elementary-events)-descriptions *U≡{Dr}, r=1,2,....s* from *(Π,U)*.

We express this goal by saying that we want to *'deprobabilise' the initial probabilistic situation by connecting it with a convenient individual integrated description of a relative form Φ*.

This, if it were achieved in fully general terms, would amount to a constructed, coherent solution to Kolmogorov's requirement of a factual interpretation of his theory of probabilities. But is this a possible aim ?

*The whole classical scientific conceptualisation involves a deterministic postulate which, in essence, asserts precisely the possibility to 'deprobabilise' any probabilistic situation*[12,13].

---

[11] In the case of a microstate this expression does not apply : the result of the (measurement)action of the view *V* is not imprinted upon the final state *of* $œ_G$, it is imprinted upon a registering device and it characterises the material elements and the conceptual and physical operations involved by V *as much as it characterises* $œ_G$ *itself*. But here we express ourselves accordingly to classical language-and-thinking.

[12] Longo,G., "Laplace, Turing et la géométrie impossible du « jeu de l'imitation » ", *Intellectica* 35, 2002.

In order to facilitate the mental representations, we place ourselves here inside the classical thinking. So we accept the probabilistic postulate. However the deterministic postulate does not include a general and effective method for also constructing the factual probability law to be asserted on a factual random phenomenon. Precisely this lacuna permits Kolmogorov's *aporia*[14]. In what follows we trace the lines of such a method.

Before entering upon this outline, let us note this. Asking *now* whether the still unknown model of a factual probability law will express what 'really' is or happens, would be but a question devoid of a possible answer and therefore entailing a state of mind paralysed by non definable *a priori* constraints. What we are entering upon is not the search of the 'discovery' of some pre-existing entity. It is the search of a method for constructing for the concept of a factual probability law, a model subject to conditions of inner logical consistency and carrying a semantic able to specify a definite and effective factual *significance* constituting an acceptable general interpretation of Kolmogorov's formal concept of a probability measure. *For this, even a model defined only in principle, would suffice*. But in fact the model proposed in what follows will arise as effectively realisable.

As for the question of 'real existence' – or of 'truth', which amounts to the same – this can be considered later, once the model has been produced. And inside *MRC* it appears that this question also can be endowed with a definite significance and with an answer, but only if it is fully relativised in a quite definite sense (Mugur-Schächter [2006] pp. 167-176).

***Construction***. *(a) Identification of the starting point*. In the case of the probabilistic game drawn from the painting $P$, the individual integrated relative description $D/G_P,P,V(El)\cup Vc\phi/$ of $P$ was available at the start, in a pre-constituted, known and actual state. It held the logical status of a primary datum. In order to extract from it a probabilistic situation, we have acted in two stages. In a first stage we have substituted to the individual integrated relative description $D/G_P,P,V(El)\cup Vc\phi/$ of $P$, a set of *100* 'local' relative descriptions $D/G_\sigma,\sigma,V(El)\cup Vc\phi/$ which permitted to reconstruct $P$ by playing a puzzle game with them. This was possible because each local form-of-colour $D/G_\sigma,\sigma,V(El)\cup Vc\phi/$ was *open* in the topological sense. It was as if not finished, not self-consistent, not closed : each border of the form-of-colour carried by one considered square $\sigma$, was source of an abstract *semantic attraction* between the form-of-colour that reached *that* border of *that* square and the form-of-colour that reached only *one* border of a *unique* other square among all the *100* squares $\sigma$ : a semantic attraction by continuity of form-of-colour when passage from the considered border of the initially chosen square, to that other privileged border of the unique other square, was examined. While this same form-of-colour that reached this same border considered on the initially chosen square, was on the contrary source of an abstract semantic repulsion by discontinuity of form-of-colour, if passage from that border to any border different from the privileged one, was considered. With respect to the goal of reconstructing the integrated individual relative description $D/G_P,P,V(El)\cup Vc\phi/$ of $P$, this sort of 'co-bordity' or 'anti-bordity' acted like a weaker substitute of the space coordinates $(x_k,y_h)$. *This* was what permitted to solve the puzzle game with the *100* local 'open' relative descriptions $D/G_\sigma,\sigma,V(El)\cup Vc\phi/$ of a form-of-colour.

But then, in a second stage, *via* the view $Vac$ of approximate *uniform* colour, we 'reduced' the *100* mutually distinct descriptions of forms-of-colour carried by the *100* squares $\sigma$, to a set $\{Dj\}\equiv\{j\}$, $j=1,2....q$ of only $q$ 'simplified' descriptions of approximate *uniform* colours, with $q \ll 100$. *This* is what brought us to a probabilistic situation, because by the passage from the *100* mutually distinct local descriptions $D/G_\sigma,\sigma,V(El)\cup Vc\phi/$ of forms-of-colour, to the set $\{Dj\}\equiv\{j\}$, $j=1,2....q$ of descriptions of approximate *uniform* colours, any connection with the individual integrated relative description $D/G_P,P, V(El)\cup Vc\phi/$ of $P$, had been effaced. It had become non perceivable in consequence of the suppression, in the new 'simplified' descriptions, of any effect of 'co-bordity' or 'anti-bordity'. With this new set of simplified relative descriptions *{Dj}≡{j}*, *j=1,2....q* only a 'probabilistic game' remained possible. And this game led to the probability space *[{Dj},$\tau_X$,{$n_P$(j)/100}]*, *j=1,2,....q* where the relative des descriptions *{Dj}≡{j}*, *j=1,2....q* acquire the status of a universe of (label-elementary-events)-descriptions.

---

[13] Modern microphysics involves the *contrary* postulate. The quantum mechanical description is built directly upon data which short-circuit classical physics. The probabilistic character of the quantum mechanical descriptions emerge as, strictly, an *empirical* 'first datum', just a primordial fact which – in consequence of the very essence of the descriptions of microstates – is non-reducible to other already *available* data, theoretical or factual. For this reason I characterise them as *primordially* probabilistic.

[14] Kolmogorov's mathematical theory reveals itself semantically insufficient when it is confronted to the quantum mechanical descriptions of microstates. But inside *MCR* it has been *extended* and reconstructed in fully relativised terms (Mugur-Schächter [2006] pp. 193-257 ; [2002] pp .256-291). The achieved result permits to incorporate a whole class of descriptions of a *primary*, basic type, which – in particular – includes the quantum mechanical descriptions of microstates. These basic descriptions, called *transferred descriptions*, constitute the very first stratum of conceptualisation, universally present beneath the classical thinking. They emerge *before* the construction of *models*. Any model can emerge only on the basis of previously elaborated transferred descriptions. By the very fact that in this work our aim is to construct a *model* for the factual probability law to be asserted on the universe of elementary events introduced by a given random phenomenon, places us inside the conceptual volume of classical thinking. This remark brings into evidence the classical character of Kolmogorov's mathematical theory of probabilities as well as of the *aporia* which flaws it *from the point of view of the classical thinking*. But, notwithstanding the fact that the conceptual domain inside which Kolmogorov's *aporia* is receivable, is a restricted domain, it is crucially important to solve this *aporia* there where it does indeed impose itself : this can endow with a precious reference permitting to understand thoroughly the specificity of quantum mechanical probabilities and to identify false problems and false assertions concerning the non classical sort of probabilities that are involved in the quantum theory. The case of the modelisation of the 'primordially' probabilistic random phenomena tied with what is called 'microstates' will be examined elsewhere (Mugur-Schächter [2008]).

This summary brings into evidence that the passage from the set of *100* mutually distinct local descriptions *D/G$_\sigma$,$\sigma$,V(El)∪Vc$\phi$/* of *forms*-of-colour, to the set *{Dj}≡{j}*, *j=1,2....q* of simplified *label*-descriptions of *uniform* approximate colours, with *q≪100*, has introduced a crucial modification, a genuine *cut* between the probability space *[{Dj},$\tau_X$,{$n_P$(j)/100}]*, *j=1,2,....q* and the individual integrated description of the painting *P*.

Indeed – in quite general terms now – when one *starts* with a probabilistic situation *(Π,U)* introduced as a primary datum, *the (label-elementary-event)-descriptions from the universe U≡{Dr}, r=1,2,....s are devoid of space-time data as well as of local 'forms' entailing 'co-bordity' or 'anti-bordity' effects. This is characteristic of a probabilistic situation*. So the problem to be solved for reaching our present constructive aim, is an *inversion* – specified in generalised terms – of the passage made in the particular case of the painting *P*, from the set of *100* mutually distinct local descriptions descriptions *D/G$_\sigma$,$\sigma$,V(El)∪Vc$\phi$/* of forms-of-colour, to the set of label-descriptions *{Dj}≡{j}*, *j=1,2....q* of approximate uniform colours, with *q≪100*.

This is not an easy inversion to be realised. In the example with the painting *P* the passage just specified was just suppressive, simplifying, destructive ; while for achieving now the researched generalising inversion, one has to prescribe a constructive method of convenient *complexification*. As it is well known, construction is much more difficult to achieve than destruction.

*(b) Qualitative model of a process of complexification of the (label-elementary-event)-descriptions {Dr}, r=1,2,....s*. Inside the probabilistic situation *(Π,U)*, *U≡{Dr}, r=1,2,....s* the (label-elementary-event)-descriptions *Dr* are generated one by one, as the experiment *Π≡[G.V]* from *(Π,U)* is repeated.

Imagine for instance repetitions of some given sort of blood analysis on samples of blood extracted from a set of *100* distinct volunteers hidden behind a screen. For each realisation of this analysis, behind the screen, one among the *100* volunteers is designated randomly *via* a drawing and he then introduces his finger inside a whole practised in the screen, thus permitting the extraction of a sample of blood. Then the result of the analysis consists in a 'label' built of answers registered on a list of pre-existing questions which filter out any features involving 'form'-generating spatial indications. This constitutes a rather creative example of a random phenomenon, in which not only the (label-elementary-event)-descriptions *Dr* do not pre-exist, but even the involved replica of the *object-entity œ$_G$ itself* – namely *any* sample of blood obtained in the conditions specified above – has to be recreated for each realisation of *Π*.

One can also think of more classical, less creative examples, in which one pre-existing object-entity œ$_G$ supposed to be endowed with permanent 'properties' independent of any realisation of *Π*, is involved in all the realisations of the considered procedure *Π*. This is so, for instance, in the paradigmatic case of throwing a dice on a table, a (label-elementary-event)-descriptions *Dr* consisting of just a number between *1* and *6* somehow indicated on the upper face of the dice when it comes to a rest position.

However, even in this case the degree of creativity in a realisation of the random procedure *Π*, is much higher than in the case of the probabilisation of the individual description of the painting *P* : each feature of the way in which any given throwing develops in time (how it evolves in the air, how it touches the table, how it changes position on the table until it stops, and so on) plays some role in the emergence of the corresponding final outcome. But in the corresponding (label-elementary-event)-descriptions *Dr*, every trace of this highly individualising evolution, is filtered out, with the unique exception of the label-value of the label-aspect *r* selected by the definition of the universe *U≡{Dr}, r=1,2,....s*.

So, each one among the possible outcomes *Dr∈U* introduced by the definition of a random phenomenon *(Π,U), U≡{Dr}, r=1,2,....s*, can be conceived of as consisting in fact of some relative description *much more complex than the obtained label-outcome Dr, consisting of a relative description of the final stage of the whole genesis of that realisation of the experiment Π which led to the obtained label-outcome Dr*.

Notice how narrowly these remarks converge with Karl Popper's famous propensity interpretation of a probabilistic situation :

> "Take for example an ordinary symmetrical pin board, so constructed that if we let a number of little balls roll down, they will (ideally) form a normal distribution curve. This curve will represent the *probability distribution* for each single experiment, with each single ball, of reaching a possible resting place. Now let us "kick" this board; say, by slightly lifting its left side. Then we also kick the propensity, and the probability distribution,.....Or let us, instead, remove *one pin*. This will alter the probability for every single experiment with every single ball, *whether or not the ball actually comes near the place from which we removed the pin*. .....we may ask: "How can the ball 'know' that a pin has been removed if it never comes near the place ? " The answer is:

the ball does not "know"; but the board as a whole "knows", and changes the probability distribution, or the *propensity*, for *every* ball; a fact that can be tested by statistical tests".

All these considerations strongly suggest what follows. Given a random phenomenon *(Π,U)*, *U≡{Dr}*, *r=1,2,....s*, a fully general method for defining a set of parcelled descriptions able to lead to the construction of an individual integrated description of a form *Φ* wherefrom the structure of the factual probability law on *U≡{Dr}*, *r=1,2,....s* be derivable, *can only stem from 'convenient' more detailed relative descriptions of the final effects of the geneses involved in the successive realisations of the procedure Π*. Indeed – apart from the universe *U≡{Dr}*, *r=1,2,....s* itself which has just been shown to have been, by construction, observationally cut off from any potentially 'corresponding' integrated description of a form *Φ* – the unique other datum that is systematically available when a random phenomenon *(Π,U)*, *U≡{Dr}*, *r=1,2,....s*, consists of the just mentioned geneses. Now, by definition, a genesis is a dynamical concept. So time must come in, amply, in the view involved in any description of a genesis. In this sense there is no symmetry with respect to the probabilising views that lead from an individual integrated description introduced as a primary datum, to a corresponding random phenomenon.

For the sake of simplicity let us suppose that the considered universe *U≡{Dr}*, *r=1,2,....s* of (label-elementary-event)-descriptions *Dr* involves only a one-aspect view *Vg*. So the *s* (label-elementary-event)-descriptions *Dr* from *U* are labelled by the *s* values *(gk)* of the unique aspect *g* (then the notation *{Dr}*, *r=1,2,....s* amounts to the condensation *[r≡(gk)]*, *k=1,2,...s*). Any other aspect which potentially might be *observed* when the outcome of an (elementary-event)-description has occurred, has been *filtered away* by the label-aspect-view *Vg*[15]. This is a manifestation of the algorithmic character of *MRC*.

Consider now the geneses assigned by the experiment *Π* to the relative descriptions *Dr∈U*. René Thom's theory of catastrophes seems particularly appropriate to be used for conceiving a model of these geneses that be useful with respect to the aim of complexifying the (label-elementary-event)-descriptions *Dr ∈U*. In the terms of Thom's theory, the set of various possible geneses involved in 'the experiment *Π*' can be conceived of as follows. Each realisation of *Π* produces a *morphogenetic* modification in a *stable substratum* that concentrates in it the stable characters of *Π*. Such a substratum necessarily exists since *Π* is said to be 'identically repeatable'. The *s* labels *r=1,2,...s* which mutually distinguish the simplified relative descriptions *Dr* from the universe *U≡{Dr}*, *r=1,2,...s*, act as *s basins of attraction* ; which means that at some stage during its evolution, the *morphogenetic* modification of the stable substratum of *Π* which is involved in a given realisation of *Π*, is captured on a sort of slope which causes it to necessarily end up with one among the *s* observable labels *Dr∈U*.

Now, a morphogenesis that realises the experiment *Π* is a *physical* entity, like also the elementary-event from *Dr∈U* by which this morphogenesis ends. So according to *MRC* (cf. the points *(6)* and *(7)*) the descriptions of these two entities do both involve some space-time background and location, and, inside the representation space of the new, complexifying view *V*, they consist of *two* 'forms' of space-time-*(gk)*-value-points where now *g=1,2...m*, *k=1,2,...w(g)*, the cardinals *m* and *w(g)* being *finite* and, given a space unit and a time unit, the values of the space and time parameters *x,y,z,t* form a *finite* space-time grid. It seems compulsory to assume that, among the new dynamical *g*-aspects (different from the label-aspect *g≡r*) which are made use of for describing the possible morphogeneses involved by *Π*, some (at least) must have imprinted some *observable traces* of their *fina*l value *gk*, upon the space-time domain that carries on it the labelling value *r* that singularises the realised (label-elementary-event)-description *Dr ∈U* by which *that* realisation of the experiment *Π* has ended. The contrary assumption seems highly unlikely. In *this* sense, the morphogenesis corresponding to the considered realisation of the experiment *Π*, finishes with *a complexified observable version* of the (label-elementary-event)-description *Dr ∈U* in the basin of attraction of which it has been captured. It finishes with a version of this *Dr∈U* which is enriched by a certain set of *observable x,y,z-gk-values of aspects g≠r* that constitute a 'form'. But *another* realisation of *Π* which ends up inside the basin of attraction labelled by the *same* value of the index *r* as in the case considered before, will in general end up with a complexified version of *Dr* which will be *different* from that from the previously considered case. So after a big number of repetitions of the experiment *Π*, every one label-description *Dr∈U*, *U≡{Dr}*, *r=1,2,...s*, will be replaced by a whole cloud of mutually distinct observable 'complexifications' of it into 'forms', each one of which emerges as the final static mark of a relative description of the whole morphogenetic process that realises the corresponding experiment *Π*. Now – contrary to what happens with the initially given label-descriptions *{Dr}*, *r=1,2,...s* themselves on which any trace of 'co-bordity' and 'anti-bordity' is absent by construction – the set of *all* these 'forms' of space-time-*(gk)*-value-points that replaces the label-descriptions *{Dr}*, *r=1,2,...s*, should *permit* to play with it a multi-dimensional puzzle game founded on semantic attractions by border-continuity or semantic repulsions by border-discontinuity among the 'forms'. Thereby the previous, probabilising passage, from a puzzle game, to a universe *{Dr}*, *r=1,2,...s* of (label-elementary-event)-descriptions that are cut from any integrated individual form, would have been reversed into a passage from the random phenomenon *(Π,U)*, *U≡{Dr}, r=1,2,....s* introduced as a primary datum, to a corresponding puzzle game.

---

[15] According to *MRC* this supposition entails no restriction whatever concerning the subsequent development. This is so in consequence of the fact that the concept of a view or an aspect-view has nothing absolute in it.

Let us now finally introduce the symbols required by the conceived model. Let $V^c$ be a *complexifying dynamical view* (**c** : complexifying) containing all the considered complexifying aspects $g \neq r$ as well as a convenient space-time frame-aspect view *V(ET)* (*MRC* the points *(6)* and *(7)*). Consider now a *given* label-description $Dr \in U$. A complexification of that *Dr* will be denoted $D_r^c(r')$, where *r'=1,2,...s'* is a **global** (undifferentiated) *aspect of complexification* endowed with *s'* values, $s' \gg 1$. (The condition $s' \gg 1$ expresses that we assume that *r'=1,2,...s'* is chosen rich enough for distinguishing the complexification of *Dr* produced by a given realisation of *Π*, from *any* other complexification of *Dr* produced by another realisation of *Π*). Then for each $Dr \in U$ we obtain a correspondence $Dr \leftrightarrow \{D_r^c(r')\}$, *r'=1,2,...s'* where the set $\{D_r^c(r')\}$, *r'=1,2,...s'* represents the cloud of mutually distinct complexifications of $Dr \in U$ produced by all the morphogeneses of *Π* which end up in the basin of attraction of *Dr* (while by construction, in consequence of the condition $s' \gg 1$, no value of *r'* is realised in $\{D_r^c(r')\}$ more than once, some values of *r'* can remain non realised inside $\{D_r^c(r')\}$).

*(c) Identification of the integrated individual description of the form Φ and numerical consequences*. Let us now specify how the individual integrated description relative to the complexifying view $V^c$, can be identified.

Imagine that the experiment *Π* from the random phenomenon *(Π,U)* is repeated a very big number *N* of times. For each realisation of *Π* the corresponding morphogenesis is registered by movies of it filmed simultaneously from several distinct view-points. These registrations are then explicitly relativised to the aspect-views from $V^c$ and finally – *via* some technique of scanning analogous to those made use of in medicine, or in architecture, or in meteorology – these relativised registrations of *Π* tied to different view-points, are unified into one multidimensional dynamical representation relative to $V^c$, of the considered realisation of *Π*. According to the model conceived above, the *ending gk*-values of this dynamical local form will determine inside the basin of attraction of some given $Dr \in U$, a multi-dimensional complexified local description $D_r^c(r')$. This local description $D_r^c(r')$, where time and dynamical aspects do no more come in, belongs to the still unknown global and a-temporal individual description of the integrated form *Φ*, which in the physical space *E* covers some still unknown surface.

While the number *N* of the repetitions of *Π* increases, progressively, the co-bordity attractions and repulsions between distinct local superficial descriptions $D_r^c(r')$, will specify for each $D_r^c(r')$ a definite location inside the multi-dimensional representation space introduced by the complexifying view $V^c$. In this way, progressively, the global integrated description of the form *Φ* will emerge inside the multi-dimensional representation space of the view $V^c$, displayed in the physical space *E* on some surface which is its integrated spatial support.

This description of the global form *Φ*, however, will not emerge *separately* for just *one* replica of it. It will be generated by a process of simultaneous and intermingled construction of an unpredictable number of replicas of this description. So a new problem arises : during the process of intermingled emergence of several replicas of the description of the global form *Φ*, one has to determine the total number $n_\Phi(r,r')$, *r=1,2,...s*, *r'=1,2,...s'* of descriptions $D_r^c(r')$ that is necessary and sufficient for building *one* replica of *Φ* and *only* one. In the case of the reconstruction of the integrated description of the painting *P* by playing puzzle, the number, *100*, of the pieces of the game was given in advance. But in the case of the construction of the integrated description of the form *Φ* the corresponding number is an *unknown* quantity. And in the present context it also is a crucial quantity, directly involved in the determination of the researched factual probability law to be asserted on the universe *U≡{Dr}, r=1,2,...s* of (label-elementary-event)-descriptions from the random phenomenon *(Π,U)*.

The solution to this problem, however, constitutes itself quasi spontaneously. Indeed, according to the theorem of big numbers, when the number *N* of the repetitions of *Π* becomes sufficiently big, it will appear with a degree of certainty which can be arbitrarily increased, that one among the simultaneously nascent replicas of the description of the integrated form *Φ* has *ceased* to offer any spatial location for supplementary descriptions $D_r^c(r')$, no matter which values of the indexes *r* and *r'* are realised in it. Then this – with a probability that can be arbitrarily increased by increasing *N* – is the first *entirely completed* replica of the description of the global form *Φ*, *achieved relatively to the complexifying view $V^c$*. On this completed replica, then, one can already count the number

$$n^c_\Phi \equiv n_\Phi[\,D_r^c(r')\,], \qquad r=1,2,...s, \qquad r'=1,2,...s', \qquad s' \gg 1 \tag{5}$$

of *all* the local complexified descriptions $D_r^c(r')$ involved in the description relative to the view $V^c$ of the integrated form $\Phi$ (the cardinal of the set $\{D_r^c(r')\}$, $r=1,2,...s$, $r'=1,2,...s'$, $s' \gg 1$).

Sooner or later, others among the intermingled simultaneously evolving replicas of the description of the global form $\Phi$ will equally manifest a stable refusal to incorporate supplementary descriptions $D_r^c(r')$, thus announcing the successive completion of also other replicas of $\Phi$. This will permit confirmation or correction of the initially found number $n^c_\Phi$ (A convenient computing program probably could easily discern this sort of emergences of a stability). So finally the key number $n^c_\Phi$ will have been established with a degree of certainty as high as one wants.

At the same time we will have completely constructed a puzzle game relative to the complexifying view $V^c$ and corresponding to the random phenomenon $(\Pi,U)$, $U \equiv \{Dr\}$, $r=1,2,....s$. (Similarly to what happened in the case of the painting $P$, each one among the $n^c_\Phi$ fragments of this puzzle carries on it one 'local' complexified description $D_r^c(r')$, which permits to play puzzle with these fragments ; furthermore, each fragment is also endowed with a set of *3* space-coordinates $x,y,z$, which permits to reconstruct the description of the global form $\Phi$ without looking at the local descriptions $D_r^c(r')$.

As for the time parameter, it has been eliminated : *the geometrisation that was our aim, is accomplished*.

So now, on the static individual description of the integrated form $\Phi$, we can also count the number of complexified versions tied with one fixed value of the index $r$, i.e. the cardinal of the set of complexifying descriptions of one given initial label-description $Dr \in U$ :

$$n^c_\Phi(r) \equiv n_\Phi[D_r^c(r')], \quad r'=1,2,...s'(r), \quad r \text{ fixed} \qquad (5)$$

And from *(4)* and *(5)* we can also estimate, for one replica of the description of the integrated form $\Phi$, the set of ratios, normalised to *1*,

$$\{ n^c_\Phi(r) / n^c_\Phi \}, \quad r=1,2,....s \qquad (6)$$

But it is *not* this set of ratios which determines the factual probability law on the random phenomenon $(\Pi,U)$, $U \equiv \{Dr\}$, $r=1,2,....s$ : in order to find this researched factual law, the data identified above have to be expressed in terms of – exclusively – the initial variable $r$.

*(d) The factual probability law on the random phenomenon $(\Pi,U)$, $U \equiv \{Dr\}$, $r=1,2,....s$.* So now, *a posteriori*, we have to make *abstraction* of all the $s'$ complexifying aspects $r'$ from the local descriptions $D_r^c(r')$. The role of these was exclusively to permit to construct from them, *via* a puzzle game, the description of the individual integrated description of the form $\Phi$. When this abstraction is operated, all the mutually distinct descriptions from the set $\{D_r^c(r')\}$, $r'=1,2,...s'$ that form the cloud of complexified description that had replaced in the puzzle game one *given* initial label-description $Dr$, are reabsorbed in that label-description $Dr$. So the number $n_\Phi(Dr)$ of realisations inside $\Phi$, of that specified label-description $Dr$, is

$$n_\Phi(Dr) = \Sigma_{r'} n_\Phi[D_r^c(r')], \quad r \text{ fixed}, \quad r'=1,2,....s' \qquad (7)$$

(in general in the sum from *(7)* some among the possible values of the index $r'$ remain non represented). So the total number, in $\Phi$, of label-descriptions $Dr \in U$, is :

$$n_{r\Phi} = \Sigma_r n_\Phi(Dr) = \Sigma_r (\Sigma_{r'} n_\Phi[D_r^c(r')]), \quad r'=1,2,....s', \quad r=1,2,....s \qquad (8)$$

On the re-expression of $\Phi$ operated above, in terms of a juxtaposition of *label*-descriptions $Dr$ from the universe introduced by the random phenomenon $(\Pi,U)$, $U \equiv \{Dr\}$, $r=1,2,....s$, let us consider now the set $\{[n_\Phi(Dr)/n_{r\Phi}]\}$ – normalised to *1* – of the *rational* ratios $[n_\Phi(Dr)/n_{r\Phi}]$, $r=1,2,.....s$. On the basis of a reasoning strictly analogous to that concerning the probabilistic game founded on the relativised parcelling of the painting $P$, we can posit that the factual and effective probability law to be asserted on the universe $U \equiv \{Dr\}$, $r=1,2,....s$ introduced by *any* given random phenomenon $(\Pi,U)$, is :

$$\{ p(Dr) \} \equiv \{ [ n_\Phi(Dr)/n_{r\Phi} ] \}, \quad r=1,2,.....s \qquad (9)$$

This achieves our constructive goal.

There does exist a model for the concept of a factual probability law.

The procedure which leads to the identification of the factual probability law to be asserted on the universe of (label-elementary-event-descriptions) from a given random phenomenon, will be called an *algorithm of semantic integration of the random phenomenon*. This algorithm defines the *significance* of the *a priori* assertion that there *exists* a factual probability law to be identified, and it also specifies the principle of an *effective* procedure for determining the structure of this factual law, by trial and error, but as rigorously as one wants.

Kolmogorov's *aporia* is solved.

Moreover, the degree of practical effectiveness of the algorithm of semantic integration of the random phenomenon seems to be largely open to various kinds of improvement. Comparisons with already existing techniques of integration, practised independently of the concept of probability, with aims concerning the fields of medicine, architecture, archaeology, meteorology, criminology, and many others no doubt, would certainly disclose the way toward optimality in the degree of effectiveness of the constructive model specified here.

And conversely also, any other method of integration, once compared to that developed here, would become itself quite fundamentally improvable in consequence of the revelation disclosed by the comparison, of the peculiar powers of precision and elucidation entailed by an explicit and rigorous insertion of a process of conceptual representation, in the general method of *relativised* conceptualisation.

Finally, let us also ask : is the description of the integrated global form $\Phi$ 'true' ? The unique answer that we are able to formulate is the following one. The mysterious relations between human thought and what we call 'physical reality', are such that usually a *logically consistent* model *built on the basis of factual data*, leads to the subsequent identification of a corresponding physical existence.

One could regard as a *problem* the fact that the probabilities from the factual law *(9)* are – by construction – exclusively rational numbers while in Kolmogorov's theory they are permitted to be any real number. But this difference, in fact, is unavoidable and also interesting. It follows from the fact that any effective factual approach is fundamentally discrete and finite, as are equally all the *observable* probabilistic features.

Let us remember Kolmogorov's assertion that « I have already expressed the view …that the basis for the applicability of the results of the mathematical theory of probability to real random phenomena must depend in some form on the *frequency concept of probability*, the unavoidable nature of which has been established by von Mises in a spirited manner…..(But) The frequency concept (of probability) which has been based on the notion of *LIMITING[16] frequency* as the number of trials increases to infinity, does not contribute anything to substantiate the applicability of the results of probability theory to real practical problems where we have always to deal with a finite number of trials ».

Whereas the formal theory of Kolmogorov is placed inside the larger abstract framework of continuous mathematics and without being rooted into a previously built factual model. In consequence of this – with respect to what it is supposed to represent – it has, by its very essence, an *approximate* and *non effective* character. We are in presence of a striking illustration of the type of false problems which stem from a surreptitious *inversion* of the roles [(to be represent*ed*) – (represent*ation*)] (G. Longo [2007], F. Bailly & G. Longo [2007], M. Mugur-Schächter [2002]). Such inversions are rather current in mathematical physics. As soon as a mathematical *language* is built in order to introduce numbers and logical security in a description of physical facts, the initial aim of mere representation tends to get forgotten and there arises an irrepressible tendency to assign hegemony to the mathematical syntax, which starts being developed *in itself*. Human mind seems to be idolatrous of mathematics. But because mathematically expressed knowledge develops by an incessant zigzag between factual descriptions and formalised descriptions, this hegemonic tendency is periodically controlled by the emergence of this or that 'problem', which leads to identify the surreptitious inversion between represented and representation wherefrom the 'problem' emerges. This is precisely what happens now in the clash between the factual concept of probability, and the mathematical one, which obliges to rediscover that factual probabilities are a basic form of our current knowledge and that consequently, an acceptable theory of probabilities has to carefully trace the characteristics of factual probabilities : *the mathematical theory of probabilities has to be reconstructed* so as to cover as rigorously and exhaustively as possible both the classical probabilistic *facts* and also those which are involved in the fundamental theory of microstates (Mugur-Schächter [2006],[2008A], [2008B]).

It is possible, however, that while achieving such a reconstruction it be found convenient to conserve a continuous mathematical framework. But this is by no means certain.

---

[16] My brackets and majuscules.

# V. CONCLUDING CONSIDERATIONS

*A posteriori*, the identification of a constructive model of the factual probability law to be asserted on a given random phenomenon, appears as unconceivable by mere use of a concept of 'elementary event' defined in the absolute classical way. With just the verbal label 'elementary event', devoid of any relativising specification of *space-time-gk*-values defining 'forms' in a definite representation space subtended by definite space-time-aspect axes, how could one have imagined some connection with a puzzle game ? In the absence, even, of any previous relation with the concept of *description*, so in the absence also of any distinction between, on the one hand, *what* is *qualified*, and on the other hand *HOW* this is qualified, relatively to what grid for qualification, structured in what a way ? Only in consequence of the powers unblocked by a rigorous insertion in *MRC*, has it been possible to solve Kolmogorov's *aporia*.

But furthermore, this insertion also opened up a clear general understanding of the roots and significance of the so fundamental but so obscure concept of probability.

In order to grasp this understanding with its whole extension, let et us start from the basic clarification introduced by *MRC* that everything which in the physical world can produce unrestricted communicable knowledge, can produce it *only* as a 'form' of space-time-aspect-values endowed with some stability, so, as a relative description which obeys the space-time frame principle (*MRC* points *(6)* and *(7)*).

As it is well known, nobody never perceives an 'object' *itself* (in the usual sense). The name of an 'object' – 'chair', 'table', etc. – points toward an *abstract* model that unifies under that name a whole set of distinct effectively perceivable relative descriptions, all assigned to a unique non-perceivable entity, but each one relative to a specific view $V$ that generates just one description from the mentioned set.

But the thickness of the screen that hides to perception and knowledge the designatum of the expression 'physical reality', has many degrees. So, in particular, it often happens that a considered 'object', which in the sense specified above *always* transgresses our capacity to perceive it 'itself', furthermore transgresses even our possibility to perceive it *at once*, by a unique act of perception, even if only from a necessarily partial point of view. Often, that which in the perceived description plays the role of object-entity $œ_G$ is *such* – with *respect* to what plays the role of view $V$ – that the description comes out parcelled, marked by division and by successivity, it comes out with the status of a 'random phenomenon'.

Nevertheless, in such a case the algorithm of semantic integration developed here permits to identify an individual integrated description of a form $\Phi$ associable to the perceived random phenomenon. Then the factual probability law derived from $\Phi$ concerning the considered random phenomenon can be regarded as a relative description of $\Phi$ in which *certain organizing space-time features that determine topological qualifications (distances, angles, etc.), got lost*. And the (label-elementary-event-descriptions) $Dr$ introduced by the considered random phenomenon, can be regarded as 'messengers' which – *via* their successive perception and their relative frequencies of emergence – offer us a sort of random and approximate 'reading' of the mentioned unknown relative description of the global form $\Phi$. But nevertheless an exact reading, in this sense that the evolving observed relative frequencies $n(Dr)/N$ reflect progressively *as exactly as one wants* the ratios $n_\Phi(Dr)/n_{r\Phi}$ from *(9)* that are realised in the constructed global relative form $\Phi$.

In short, *inside MRC a probabilistic description acquires a constructible intelligible referent $\Phi$*.

_________

**BIBLIOGRAPHY**

Mugur Schächter, M., :

- [1991], "Spacetime Quantum Probabilities I:..... ", *Founds. of Phys.*, Vol. 21, pp. 1387-1449 ;

- [1992A], "Toward a Factually Induced Space-Time Quantum Logic", *Founds. of Phys.*, Vol. 22, pp. 963-994 ;

- [1992B], "Quantum Probabilities, Operators of State Preparation, and the Principle of Superposition", *Int. J. of Theoretical Phys.,* Vol. 31, No.9,1715-1751 ;

- [1993], "From Quantum Mechanics to Universal Structure of Conceptualization and Feedback on Quantum Mechanics", *Founds. of Phys.*, Vol. 23, No 1, pp. 37-122 ;

- [2008], "L'infra-mécanique quantique", 167 pages, to be published.

Kolmogorov A.N., :

- [1933], "Grundbegriffe der Wahrscheinlichkeitrechnung", *Ergebnisse der Mathematik* ;

- [1950]., "Foundations of the Theory of Probabilities", *Chelsea Publishing Company*, translation of the original monography from 1933.

Shannon, E.C., [1948], "The mathematical Theory of Communication", *Bell Syst., Techn. Journ.,* 27, 379-423 ; 623-656.

Khinchin, A.I., [1957], "Mathematical Foundations of Information Theory", *Dover Publications*.

Solomonoff, R. J., [1957], "An inductive inference machine", *IRE National Record*, 5 (quoted in :
Segal, J., [2003], "Le zéro et le un", Syllepse).

Kolmogorov A.N., [1963], in *Sankhya*, (quoted in :
Segal, J., [2003], "Le zéro et le un", Syllepse).

Kolmogorov, N.A., [1983], "Combinatorial foundations of information theory and the calculus of probabilities", *Russia Mathematical Surveys*, 38, pp. 29-40.

Mugur-Schächter, M., :

- [1984] "Esquisse d'une représentation générale et formalisée des descriptions et le statut descriptionnel de la mécanique quantique", *Epistemological Letters*, Lausanne, cahier 36, pp. 1-67;

- [1992C] "Spacetime Quantum Probabilities II : Relativized Descriptions and Popperian Propensities", *Founds. of Phys*., Vol. 22, pp. 269-303 ;

- [1995] "Une méthode de conceptualisation relativisée... ", *Revue Int. de Systémique*, Vol. 9 ;

- [2002A] "Objectivity and Descriptional Relativities", *Foundations of Science* 7, 73-180 ;

- [2002B] "Quantum Mechanics versus Relativised Conceptualisation" in *Quantum Mechanics, Mathematics, Cognition and Action*, M. Mugur-Schächter & A. Van der Merwe, pp. 109-307 ;

- [2006] "Sur le tissage des connaissances", Hermès Science Publishing-Lavoisier, 316p.

Longo,G., [2007], "Laplace, Turing and the 'imitation game' impossible geometry: randomness, determinism and programs in Turing's test", in Epstein, R., Roberts, G., & Beber, G. (Eds.). The Turing Test Sourcebook. Springer.

Bailly, F. & Longo, G., [2007], "Randomness and Determination in the interplay between the Continuum and the Discrete", in Special issue: Mathematical Structures in Computer Science 17(2), pp. 289-307.

Mugur-Schächter, M., [2002C], "En marge de l'article de Giuseppe Longo sur Laplace, Turing et la géométrie impossible du 'jeu d'imitation' ", *Intellectica* 2002/2,pp. 163-174.